\newcommand{\comment}[1]{}

\def \rightdownarrow
 {\kern.3em
 \rule[.5ex]{.15mm}{2ex}
 {\mbox{$\kern-0.1em{\longrightarrow}$}}
 }
\def\lessim{\mathrel {\vcenter {\baselineskip 0pt \kern 0pt
\hbox{$<$} \kern 0pt \hbox{$\sim$} }}}
\def\gessim{\mathrel {\vcenter {\baselineskip 0pt \kern 0pt
\hbox{$>$} \kern 0pt \hbox{$\sim$} }}}
\newcommand{\beq}{\begin{equation}}
\newcommand{\eeq}{\end{equation}}
\newcommand{\bear}{\begin{array}}
\newcommand{\ear}{\end{array}}
\newcommand{\bet}{\begin{tabular}}
\newcommand{\eet}{\end{tabular}}
\newcommand{\beqn}{\begin{eqnarray}}
\newcommand{\eeqn}{\end{eqnarray}}

\newcommand{\bfh}{\begin{figure}[h]}
\newcommand{\efh}{\end{figure}[h]}
\newcommand{\ee}{\mbox{$e^{+}e^{-}$}}

\newcommand{\tev}{\ensuremath{\mathrm{Te\kern -0.1em V}}}
\newcommand{\gev}{\ensuremath{\mathrm{Ge\kern -0.1em V}}}	
\newcommand{\mev}{\ensuremath{\mathrm{Me\kern -0.1em V}}}	
\newcommand{\kev}{\ensuremath{\mathrm{ke\kern -0.1em V}}}	
\newcommand{\massmev}{\mbox{\mev/$c^2$}}			

\newcommand{\bd}{\ensuremath{B^{0}}}				
\newcommand{\bs}{\ensuremath{B^{0}_s}}				
\newcommand{\bu}{\ensuremath{B^{+}}}				

\newcommand{\abs}{\ensuremath{\overline{B}^{0}_s}}		

\newcommand{\bskk}{\ensuremath{\bs \to  K^+ K^-}}

\newcommand{\Bd}{$B^{0}$}

\newcommand{\Bu}{$B^{+}$}
\newcommand{\Bs}{$B_{s}^{0}$}
\newcommand{\Lb}{$\Lambda_{b}^{0}$}

\newcommand{\Bdmumu}{\ensuremath{\bd \to \mu^{+}\mu^{-}}}
\newcommand{\Bsmumu}{\ensuremath{\bs \to \mu^{+}\mu^{-}}}

\newcommand{\lumifb}{\mbox{fb$^{-1}$}}				

\newcommand{\Dzero}{\ensuremath{\mathrm{D}\emptyset}}
\newcommand{\Sbp}{\ensuremath{\Sigma_{b}^{+}}}
\newcommand{\Sbm}{\ensuremath{\Sigma_{b}^{-}}}
\newcommand{\Sbpst}{\ensuremath{\Sigma_{b}^{*+}}}
\newcommand{\Sbmst}{\ensuremath{\Sigma_{b}^{*-}}}
\newcommand{\Omegab}{\ensuremath{\Omega_{b}^{-}}}
\newcommand{\betas}{\ensuremath{\beta_{s}}}
\newcommand{\bsphiphi}{\ensuremath{\bs \to \phi\phi}}
\newcommand{\Xib}{\ensuremath{\Xi_{b}^{-}}}
\newcommand{\bsbar}{\ensuremath{\overline{B}^0_s}}

\documentclass{PoS}

\title{Flavour physics at the Tevatron}

\ShortTitle{Flavour physics at the Tevatron}

\author{\speaker{Giovanni PUNZI}%
        \thanks{ }\\
       Universita' di Pisa and INFN, Largo B. Pontecorvo 3,  56127 Pisa, Italy\\
       E-mail: \email{giovanni.punzi@pi.infn.it}}

\abstract{The Tevatron heavy flavor physics program is in full swing. The rapid increase in the size of data samples is allowing significant improvements of previous results, and opens the doors to new possibilities. A further doubling of the current integrated luminosity is expected in the next couple of years. This report summarizes the main current results and future prospects.}

\FullConference{European Physical Society Europhysics Conference on High Energy Physics\\
		 July 16-22, 2009\\
		 Krak\'{o}w, Poland}

\begin{document}


\section{Properties of Heavy-Flavored hadrons}

The rich production of heavy hadrons of all kinds, together with 
limited event pile-up and high-quality tracking detectors puts the 
Tevatron in a very favorable position for many experimental studies of heavy--flavored hadrons. 
Some topics are even exclusive to Tevatron. One example in this category is the properties of bottom--flavored baryons. Just few years ago, the \Lb\ was the only bottom baryon for which some significant 
experimental information was available. In the past couple of years, knowledge of 
$b$-baryons has expanded enormously thanks to Tevatron data. In 
year 2007, CDF observed the strongly--decaying
\Sbp\ ($uub$) and \Sbm\ ($ddb$) baryons\footnote{N.B. the two oppositely--charged states are not a particle--antiparticle pair.}, together with their orbitally excited 
($J=3/2$) partners \Sbpst , \Sbmst , and measured their masses with a precision at the
MeV level\cite{Sigmab}. 
That same year, both \Dzero\ and CDF observed the charged \Xib\  
baryon ($usb$), making the \Lb\ not anymore the only known example of weakly--decaying bottom baryon~\cite{CDFXib,D0Xib}. 
Finally, last year  \Dzero\ announced the observation of the doubly--strange \Omegab\ baryon ($ssb$), the heaviest 
fundamental state of a single--bottomed baryon, recently followed by a similar result from CDF. 
The masses of heavy--flavored baryons are more than just a "zoological" curiosity: they allow to test 
our understanding of the dynamics involved in the structure of 
hadrons. The availability of a whole spectrum of measurements from the Tevatron is particularly 
important in this respect, as most models are much better at predicting differences of masses or other overall relationships than individual masses. 
Here we discuss the most recent results, that are about the weakly--decaying baryons \Xib\ and 
\Omegab .
Both states are easily triggered and reconstructed at both CDF and 
\Dzero\ through their decays into $J/\psi \to \mu^{+}\mu^{-}$ : $\Omegab \to J/\psi \Omega^{-}$ , $\Omega^{-}\to 
\Lambda^0 K^{-}$ and $\Xib \to J/\psi \Xi^{-}$ , $\Xi^{-}\to 
\Lambda^0 \pi^{-}$. 
The presence of multiple long--lived particles makes for a 
three--vertex, 5--track configuration. 
In the case of \Dzero , a 
special--purpose tracking code has been used to improve efficiency 
for reconstruction of tracks with significant impact parameters to the beamline. 

\begin{figure}
\mbox{\includegraphics[width=2in,height=1.5in]{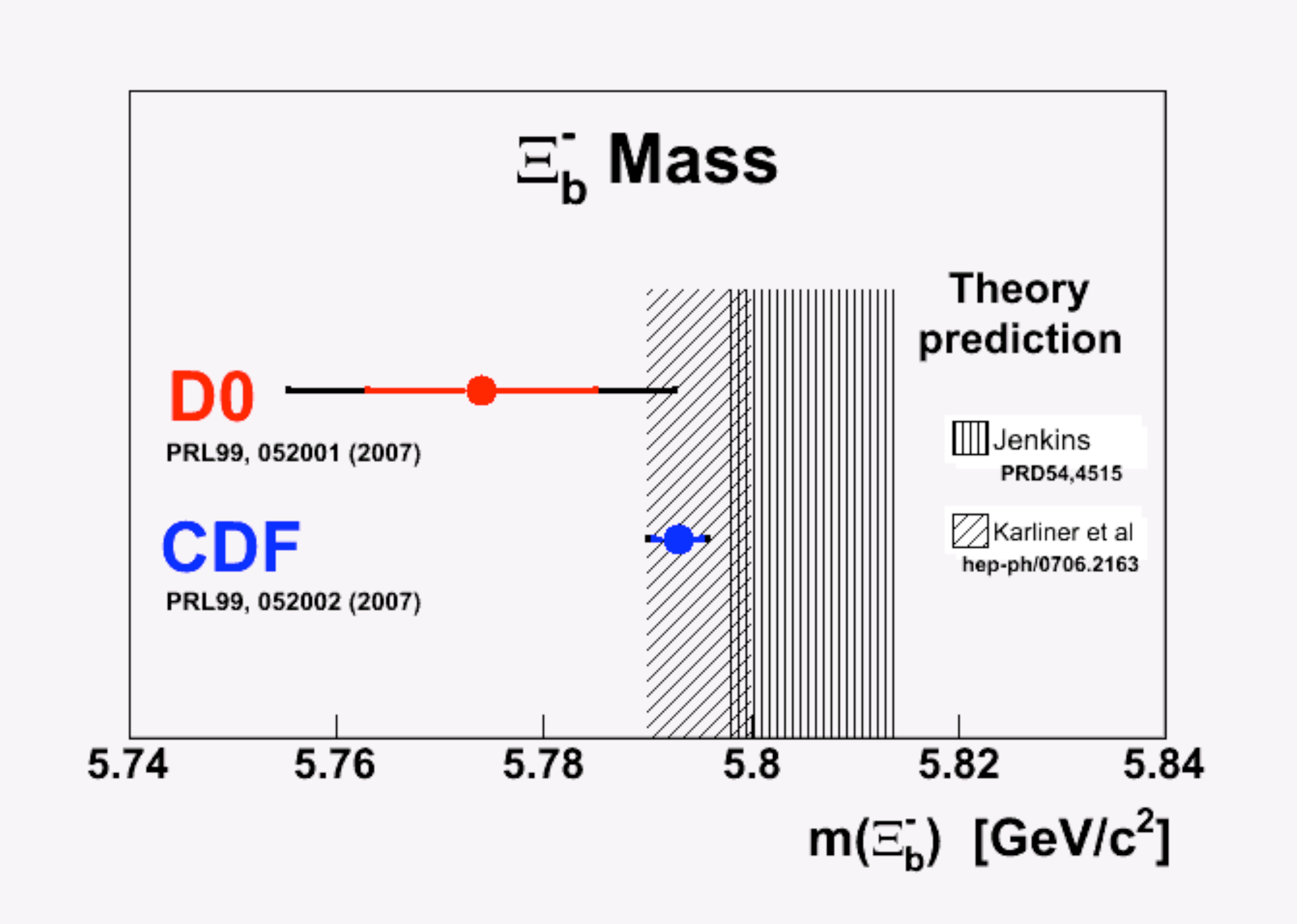}}
\mbox{\includegraphics[clip, trim=0in 0.4in 0in 0in ,width=4in,height=2.2in]{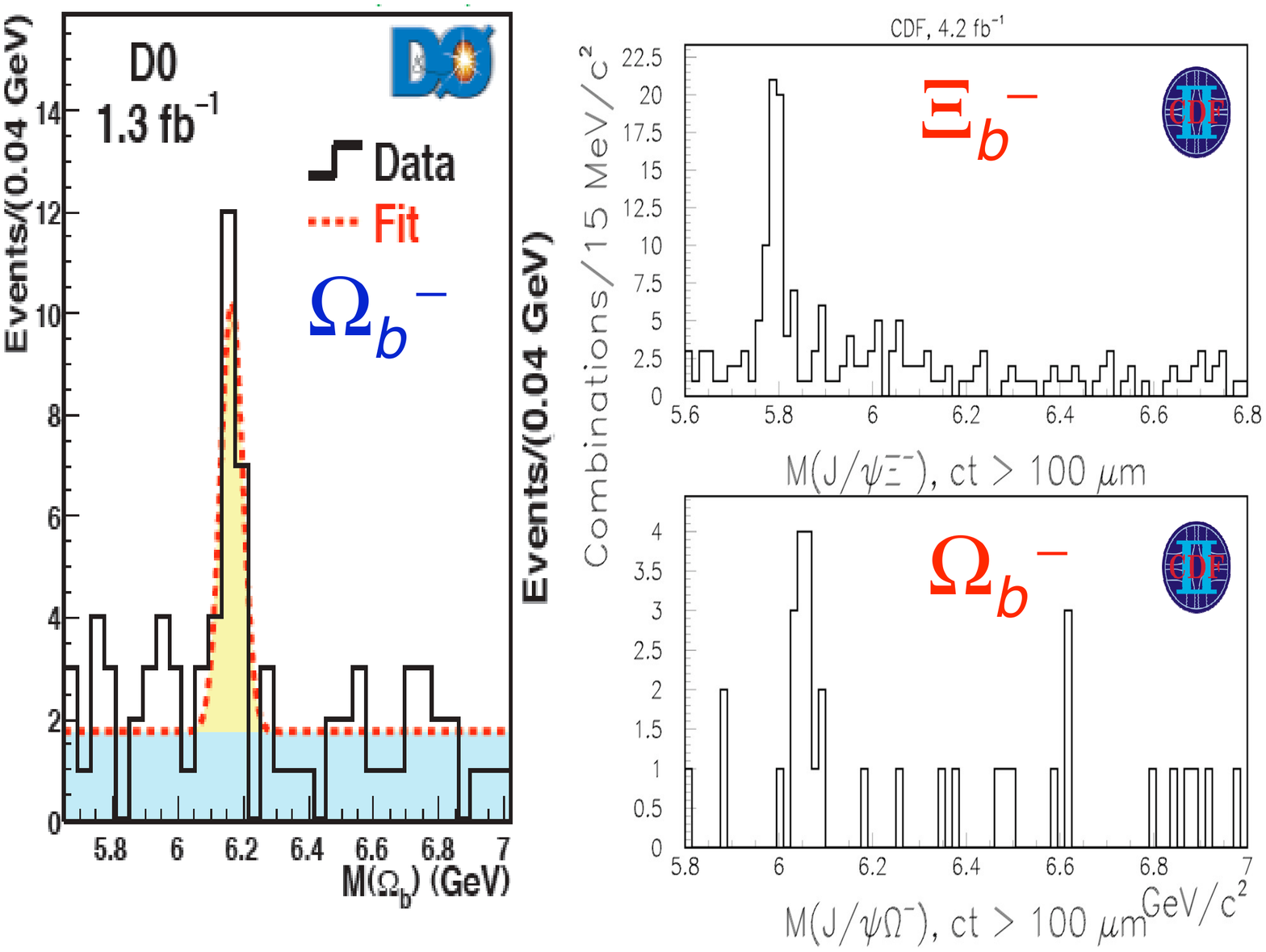}}
\caption{\Xib\ masses (left) and \Omegab\ ,\Xib\ signals (right)}
\label{fig:xib}
\end{figure}

The initial measurements of the \Xib\ baryon mass are shown in Fig.~\ref{fig:xib}. They are in good agreement with each other and with predictions, although the CDF measurement is significantly more precise, matching the resolution of the most precise predictions available.
CDF has recently performed an update with a 4.2\lumifb\ sample, performing a global analysis aimed at both the \Xib\ and \Omegab\ baryons, and including several other $B$ meson channels of similar decay topology as controls~\cite{omegaCDF}. The result: $M(\Xib) = 5790.9\pm 2.6 \pm 0.8\massmev$ is compatible with the previous measurement and supersedes it. For the \Omegab , CDF finds $M(\Omegab)=6054.4 \pm 6.8\pm 0.9 \massmev $. This level of precision allows to actually discriminate between different models: for instance, the new masses measured by CDF are in close agreement with recent lattice~\cite{latticemasses} or color--hyperfine--splitting calculations~\cite{baryonmasses}, with a resolution that appears to allow even a discrimination between different forms of the potential. However, there is a $6\sigma$ disagreement between the CDF measurement of $M(\Omegab)$ and the value previously measured by \Dzero : $M(\Omegab)=6165 \pm 10 \pm 13  \massmev $. In addition, the production rate is also not very well matched, although both experiments claim signal significances above $5\sigma$, (see~\cite{omegaCDF,omegaD0} for more details). This puzzle must be solved before definitive conclusions can be drawn, and this will likely not happen without the help of further experimental data.


Both CDF and D0 find the decay time distribution for the \Omegab\ to be compatible with a weakly-decaying $b$-hadron. CDF also quotes explicit lifetime measurements.  The $\Xib$ lifetime is measured exclusively for the first time, yielding: $\tau(\Xib) = 1.56^{+0.27}_{-0.25}\pm 0.02$ ps; the result is consistent with the average lifetime obtained at LEP for a mix of $\Xi_b^0$, \Xib , and $\Lambda_b$ baryons~\cite{PDG}. For the \Omegab\ this is the first lifetime measurement: $\tau(\Omegab) = 1.13^{+0.53}_{-0.40}\pm 0.02$ ps, again compatible with a weakly--decaying $b$ hadron, and slightly favoring a shorter lifetime than the \bd .
In addition to these unique measurements on bottom baryons, many other measurements of bottom hadron lifetimes have been performed at the Tevatron. The lifetime of the unique doubly heavy--flavored meson $B_c$ has been measured with 10\% resolution~\cite{Bclifetime}, and the current world averages of  \Lb\ and \Bs\ lifetimes are dominated by the Tevatron measurements performed with 1\lumifb ~\cite{Fernandez,HFAG}. All such measurements are expected to improve with the large samples now available and not yet analyzed ($\approx 5 \lumifb $).
Even the \Bd\ and \Bu\ lifetimes, for which large samples are already available from  \ee\ 
experiments, will be further improved and become dominated by the Tevatron measurements. 


Progress has also been made in the field of exotic heavy hadrons. CDF 
has the largest existing sample of X(3872), the first and most studied exotic state, allowing precision measurements 
aimed at understanding the nature of this still mysterious particle. 
In the past these have allowed to strongly constrain the possible $J^{PC}$ 
assignments~\cite{X3872angular}. 
The latest analysis of a 2.4\lumifb\ 
sample, corresponding to $\simeq 6,000$ signal events, yields the most precise mass measurement: $M(X(3872))= 3871.61\pm 0.16\pm 0.19\massmev$, and allows a 
detailed analysis of its lineshape, excluding the possibility of a 
two--component system with a separation greater than 3.6 \massmev\ in the assumption of equal production~\cite{X3872}. 
The measured X(3872) mass is lower than 
$M(D^{0})+M(D^{\ast 0})$, still allowing the possibility of interpretation as a loosely bound ``molecular''
system. However, the masses are compatible within uncertainty, the difference being just $0.19 \pm 0.43 \massmev$, with the uncertainty being dominated (somewhat paradoxically) by the PDG uncertainty on the 
masses of the common $D^0$ mesons. 
The above observations favor molecular models over tetraquark 
models; however, the observed large production cross section of the X(3872) at the Tevatron seems very difficult 
to reconcile with a molecular model~\cite{Piccinini}, so the issue is far 
from being settled, and the true nature of the X(3872) remains a fascinating puzzle.

The study of further states may help understanding what is going on. 
The latest appearance on the scene is the Y(4140), a $3.8 \sigma$ excess observed by CDF in the $J/\psi\phi$ mass spectrum with mass $M(Y(4140)=4143.0 \pm 2.9 \pm 1.2 \massmev $~\cite{Y4140}. 
This has been reconstructed in a sample of $75 \bd \to J/\psi \phi K^{+}$ decays
(also the largest existing sample of this mode) obtained from a 2.7\lumifb\ sample. 
The nature of this signal is still unclear and will benefit from the addition of further data.

\section{Rare Decays}
\begin{figure}
\includegraphics[clip, trim= 1.2in 1.5in 1.5in 2in, width=\textwidth]{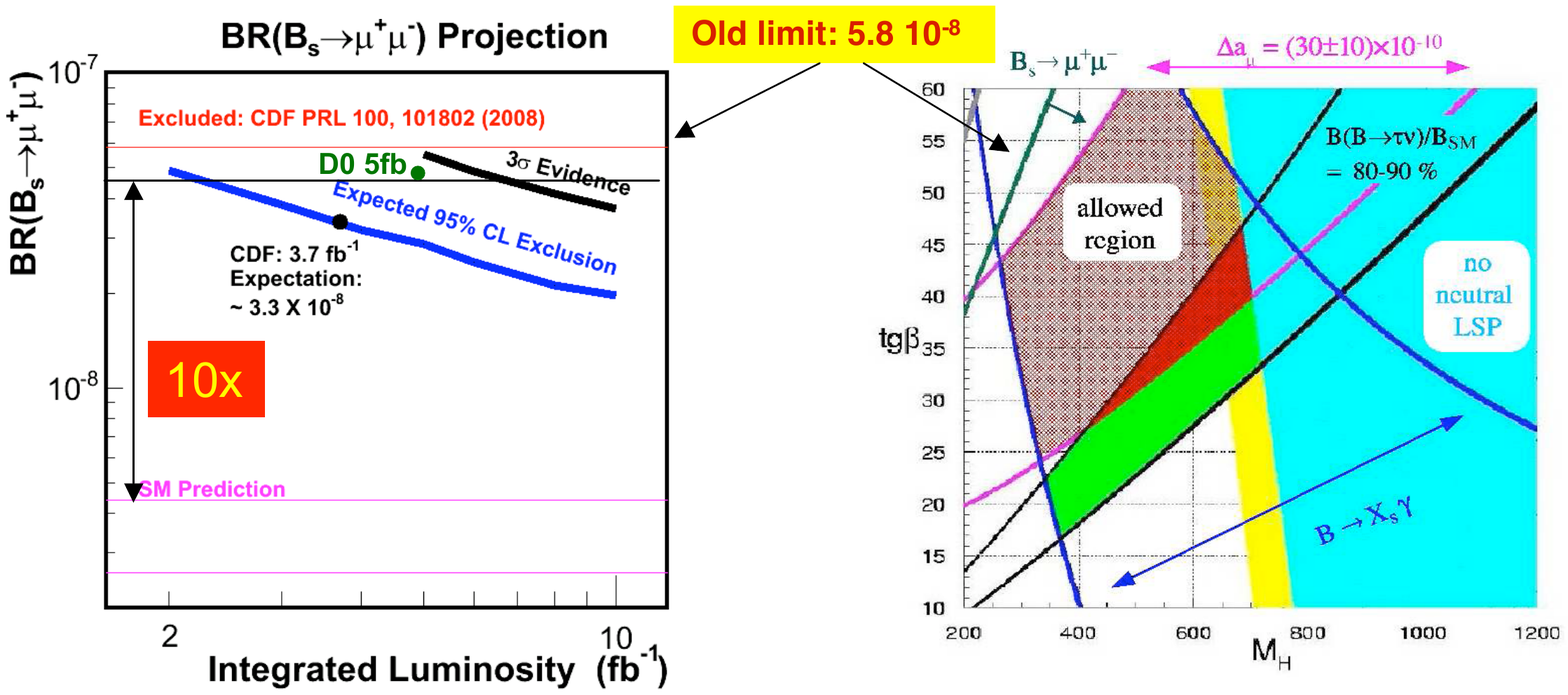}
\caption{Current results (left) and constraints imposed by some measurements on the MSSM model (heavy squarks)~\cite{Isidori:2008qp, Isidori:2007jw}(right)}
\label{fig:bsmumu}
\end{figure}

The large B and D production rate and the strong 
background rejection capabilities of the two detectors make the Tevatron an ideal place to look for rare heavy flavor decays. 
Several rare modes have been studied, but the channels attracting the greatest interest are 
\Bsmumu\ and \Bdmumu . These  Flavor Changing Neutral Current 
processes have small branching fractions in the Standard Model, but may receive contributions from a large variety of BSM processes. The latest and most precise predictions of branching fractions in the Standard Model are  $BR(\Bsmumu)= (3.6\pm 0.3) \times 10^{-9}$ and $BR(\Bdmumu)=  (1.1\pm 0.1)  \times 10^{-10}$~\cite{Buras:2009us}. It is worth noting that the ratio between the \bd\ and the \bs\ modes is not necessarily preserved in SUSY models, and the \bd\ might even be larger than the \bs\ (see for instance~\cite{Dedes:2008iw}), and it is therefore important to study both separately.

Both CDF and \Dzero\ have now analyzed significantly larger samples since their last publications~\cite{mumuold}, but the final results are still ``blind''. The new \Dzero\ analysis uses about 4\lumifb\ , divided into three different run ranges, that are analyzed independently, and then combined to produce a single limit with the CLs algorithm~\cite{talk:Ripp-Baudot}. A Boosted Decision Tree technique is used  to optimize the selection, by minimizing the expected upper limit under the assumption of no signal, resulting in an expected limit of $4.3(5.3)\times10^{-8}$ at 90(95)\% C.L.

The new CDF analysis is based on 3.7\lumifb\ and makes use of an Artificial Neural Network (ANN, or NN) to optimize the selection. Data are not divided into periods, but are instead divided in bins of NN output, to exploit differences in signal/background ratio to increase sensitivity. The new analysis has an increased geometrical acceptance with respect to the past, owing to the addition of some further detector regions. The optimization procedure is otherwise the same as done by \Dzero . The new expected limit for the \Bsmumu\ is $4.3\times 10^{-8}$, almost a factor 2 better than previous results\footnote{CDF has since then unblinded their result, obtaining: $BR(\Bsmumu)<4.3\times 10^{-8}$ and $BR(\Bdmumu)<7.6 \times 10^{-9}$ at 95\% C.L. 
The expected limit for \Bsmumu\ went down to $3.3\times 10^{-8}$, and the slight excess over prediction is quantified as a $0.73\sigma$ effect~\cite{BsmumuCDF}.}. 

It is interesting to note that in this update CDF and \Dzero\ have an expectation for about one SM event each in their analyzed samples, although the background conditions prevent detecting it. The two experiments are now really closing in towards the Standard Model expectation, cutting into the final factor of ten, and will have significant impact in constraining the possibilities for the existence of new physics (see Fig.~\ref{fig:bsmumu} for an example). Work is ongoing to include all data up to 5\lumifb\ and implement some further improvements to push the sensitivity of the analysis even further. A Tevatron average is likely to be very interesting at this stage.

Several other results on rare modes are being produced in parallel with \Bdmumu\ and \Bsmumu .  Using data from the impact parameter trigger rather than the dimuon trigger, CDF has recently published results from the search of $\bd , \bs \to \ee, e^{+}\mu^{-}$  that set the tightest existing limits on their branching fractions, and excluded the existence of leptoquark states with masses below $\simeq$50~TeV~\cite{Bemupaper}. From the same trigger selection, tight limits on charm FCNC decays are also obtained~\cite{CDF9226}.
Rare modes of the type $\mathrm{B} \to X_s \ell^+\ell^-$ have already been observed at CDF with 1\lumifb ~\cite{Aaltonen:2008xf}, and an updated analysis with the full statistics is ongoing\footnote{The analysis has been released at HCP 2009 and includes the first observation of the $\bs \to \phi \mu^+\mu^-$ mode.}. These channels are particularly interesting in view of the recent hints for anomalies in the $A_{FB}$ distributions seen in \ee\ experiments~\cite{BelleXll,Bevan}. A confirmation of these anomalies might signal the presence of a 4th generation or other BSM phenomena.

\section{\bs\ mixing parameters}

The \bs --- \bsbar\ system is an excellent lab for testing our understanding of SM physics and look for what may lie beyond that. Most conceivable new physics models have some impact on \bs\ oscillations, and the measurements of oscillation parameters work coherently together with the FCNC branching fraction measurements discussed in previous section to provide powerful constraints on BSM physics (see for instance~\cite{Burastalk, Buras:2009us} and references therein).

Experimentally observable quantities are:

\begin{eqnarray*}
\Delta m_s = M_H - M_L , 
\Gamma_s = \frac{\Gamma _L+\Gamma _H}{2} , 
\Delta\Gamma_s = \Gamma _L - \Gamma _H , 
a^s_{fs} = \frac{\Gamma (\bar{B}_s(t) \to f) - \Gamma (B_s(t) \to \bar{f})}{\Gamma (\bar{B}_s(t)\to f) + \Gamma (B_s(t) \to \bar{f}) }
\end{eqnarray*}

They can be conveniently parametrized as follows:
\begin{eqnarray}
\Delta m_s = \Delta m_s^{\mathrm{SM}} |\Delta_s|\\
\Delta\Gamma_s = \Delta\Gamma_s^{\mathrm{SM}} \cos (\phi_s^{\mathrm{SM}} + \phi_s^\Delta)\label{eq:cosphis}\\
\frac{\Delta\Gamma_s}{\Delta m_s} = \left( \frac{\Delta\Gamma_s}{\Delta m_s}\right)^{\mathrm{SM}} \frac{\cos (\phi_s^{\mathrm{SM}} + \phi_s^\Delta)} {|\Delta_s |}\\ 
a^s_{fs} = \frac{\Delta\Gamma_s}{\Delta m_s} \tan(\phi_s^{\mathrm{SM}} + \phi_s^\Delta)
\end{eqnarray}

where $|\Delta_s|$ and $\phi_s^\Delta$ parameterize modulus and phase of possible non--SM contributions, and $\phi_s^{\mathrm{SM}}$ is small ($\sin(\phi_s^{\mathrm{SM}}) \simeq 0.004$).
In addition, we use
\begin{equation}
\phi_s^{J/\psi \phi} \equiv -2\beta_s^{J/\psi \phi} \equiv \phi_s^{\mathrm{SM}} + \phi_s^\Delta
\end{equation}

(In the following, will omit the ``$J/\psi \phi$" label for better readability).

The current knowledge of the \bs\ mixing parameters is dominated by the Tevatron measurements, and will stay that way until the LHCb will have collected and analyzed a sizeable sample. 
The best known parameter is the oscillation frequency: $\Delta m_s = 17.77\pm 0.10\pm 0.07 \mathrm{ps}^{-1}$. The experimental uncertainty is below the level of the theory uncertainty, and at the moment there is no obvious motivation for further improving it. The width difference is measured with limited precision: $\Delta\Gamma_s = 0.062 ^{+0.034}_{-0.037}$ ps$^{-1}$~\cite{PDG}. This value is compatible with the latest predictions: $\Delta\Gamma_s = 0.088 \pm 0.017 \mathrm{ps}^{-1}$~\cite{Lenz:2006hd}, but it is also compatible with zero. However, this world-average was derived under the specific assumption of no CPV in the \bs\ mixing, which is related to the mixing phase $\phi_s$. For this reason, and because of Eq.~\ref{eq:cosphis}, a simultaneous determination of $\Delta\Gamma_s $ and $\phi_s \equiv -2\beta_s$ is a better option. The phase $\phi_s$ is currently known only thanks Tevatron measurements of the time and angular distribution of  $\bs\to J/\psi\phi$,  and still has a pretty large uncertainty. The asymmetry $a_{fs}$ is hard to measure accurately. The only existing data is from the recent measurement by the \Dzero\ collaboration using semileptonic decays ($A^s_{SL}$), yielding the result~\cite{Abazov:2009wg}: 
\begin{equation}
A^s_{SL} = \frac{N(\abs\to\ell^+\nu_\ell X) - N(\bs\to\ell^-\overline{\nu}_\ell X)}{N(\abs\to\ell^+\nu_\ell X) + N(\bs\to\ell^-\overline{\nu}_\ell X)} = -0.0017\pm0.0091^{+0.0012}_{-0.0023}
\end{equation}
The \Dzero\ capability of regularly inverting the B field direction contributes to keep the systematic uncertainty low.
Although the best existing measurement, this is still compatible with zero, and far from testing the SM prediction of $2\times 10^{-5}$.

The measurement of  ($\Delta\Gamma_s $, $\mathrm{\phi_s}$) parameters is based on an unbinned fit of the decay time and angular distribution of the $\bs \to J/\psi \phi$ decay. Both CDF and \Dzero\ have performed  this measurement with 2.8\lumifb ; the reader is referred to other contributions in this conference for details~\cite{betas_CDF,betas_D0}. Here we report for the first time the Tevatron combination of  ($\Delta\Gamma_s $, $\mathrm{\phi_s}$) . 
This combination was performed by the Tevatron B averaging Working Group and includes CDF and \Dzero\ results with 2.8\lumifb\ of data~\cite{TEVBWG}. In order to produce this average, the CDF and \Dzero\ collaborations have worked together to adapt their initial measurements to make them compatible, making their procedures much more similar, to ensure the combined result is meaningful. As a consequence, the \Dzero\ result has changed slightly~\cite{betas_D0}. 

\begin{figure}
\begin{center}
\includegraphics[width=4.5 in]{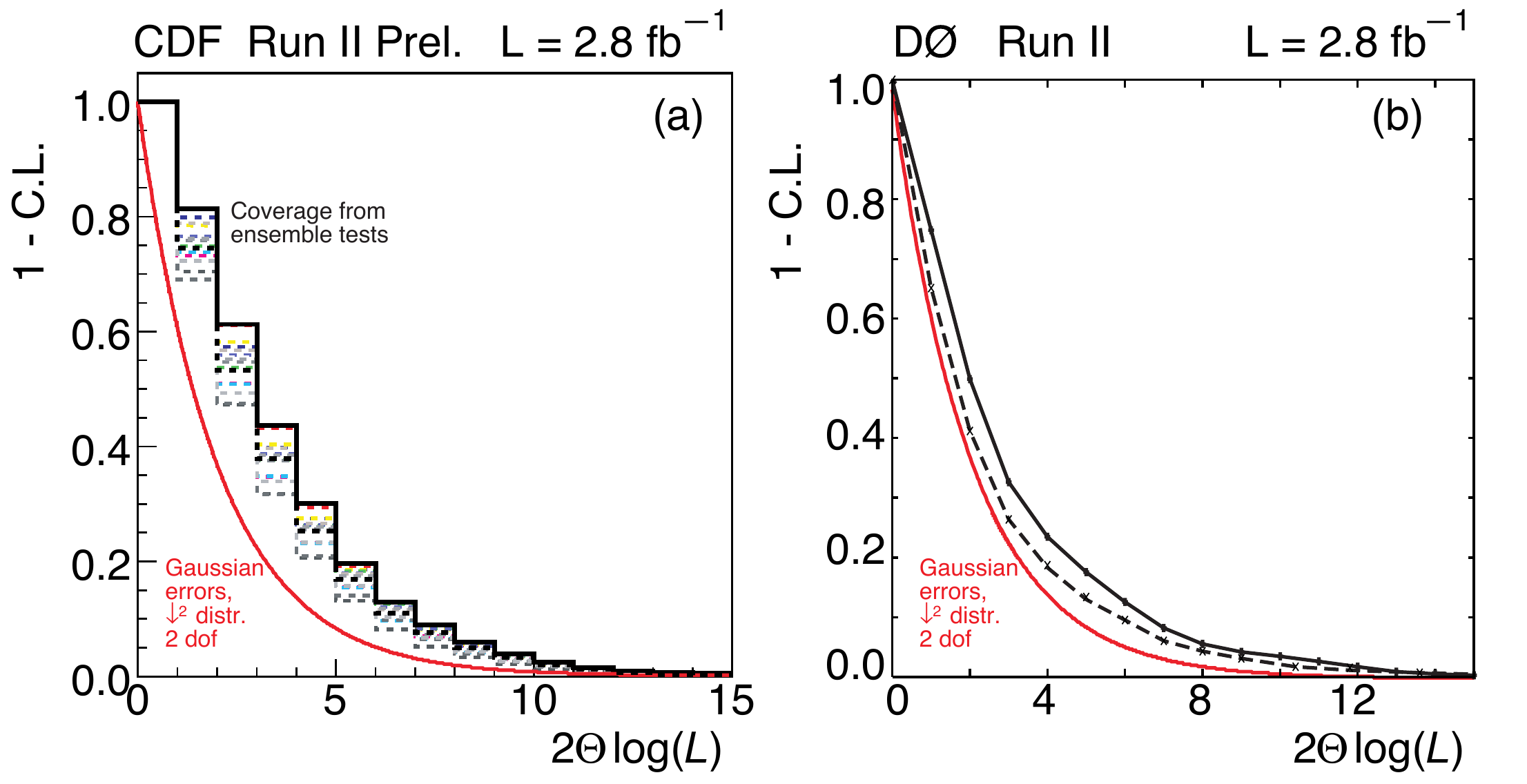}
\caption{Observed distributions of the profile likelihood ratio quantity for ($\Delta\Gamma_s $, $\mathrm{\phi_s}$) measurement, compared to the nominal asymptotic chi-square distribution.}
\label{fig:pLRdistr}
\end{center}
\end{figure}

Both experiments now use a common method for the extraction of confidence contours in the  ($\Delta\Gamma_s $, $\mathrm{\phi_s}$)  plane. Due to the critical impact that this measurement may have in revealing new physics, special care has been taken in avoiding approximations and staying on the conservative side if necessary. A complete frequentist construction with exact inclusion of systematic uncertainties is used. A profile likelihood function $L_{prof}(\Delta\Gamma_s ,\mathrm{\phi_s})$ is built for the two parameters of interest, by maximizing the likelihood over all other unknown parameters (physical parameters, like strong phases, and systematic parameters), and the ratio is taken with the value of the function at the mimum $LR_{prof}$. The use of external constraints in the likelihood is minimized, and made equal between the two experiments. Due to limited statistics and strong non-gaussian effects, the usual assumption that the probability distribution of $LR_{prof}$ has the asymptotical chi--squared distribution is not a good approximation. The  $LR_{prof}$ is therefore used just as the ordering function in a full Neyman construction of a multidimensional confidence region in the complete parameter space, which is then projected over the 2D space of the parameters of interests  ($\Delta\Gamma_s $, $\mathrm{\phi_s}$) . This procedure has been proved to produce confidence region with close-to-optimal perfomance in terms of sensitivity~\cite{Oxford}, and is critical to avoid an unnecessary weakening of the limits when the large parameter space ($\simeq 30$ dimensions) is projected over the 2D space. 
The projection is performed by Monte Carlo generation of $LR_{prof}$ distributions for a number of points randomly sampled within the nuisance parameter space, and choosing the distribution with the longest tail ("worst-case") as the basis for the construction of the confidence region. It is interesting to note that the deviation of the distributions obtained numerically from the nominal behavior is significant and cannot be ignored (Fig.~\ref{fig:pLRdistr}). In order to avoid divergence of the size of the confidence regions, the allowed range of variation of the nuisance parameters was loosely bounded without compromising the coverage of the procedure, by exploiting the method of ref.~\cite{BB}.
Finally, the 2D confidence regions obtained in this way are combined with the usual HFAG method~\cite{HFAG}. 

\begin{figure}
\includegraphics[width=3.5in]{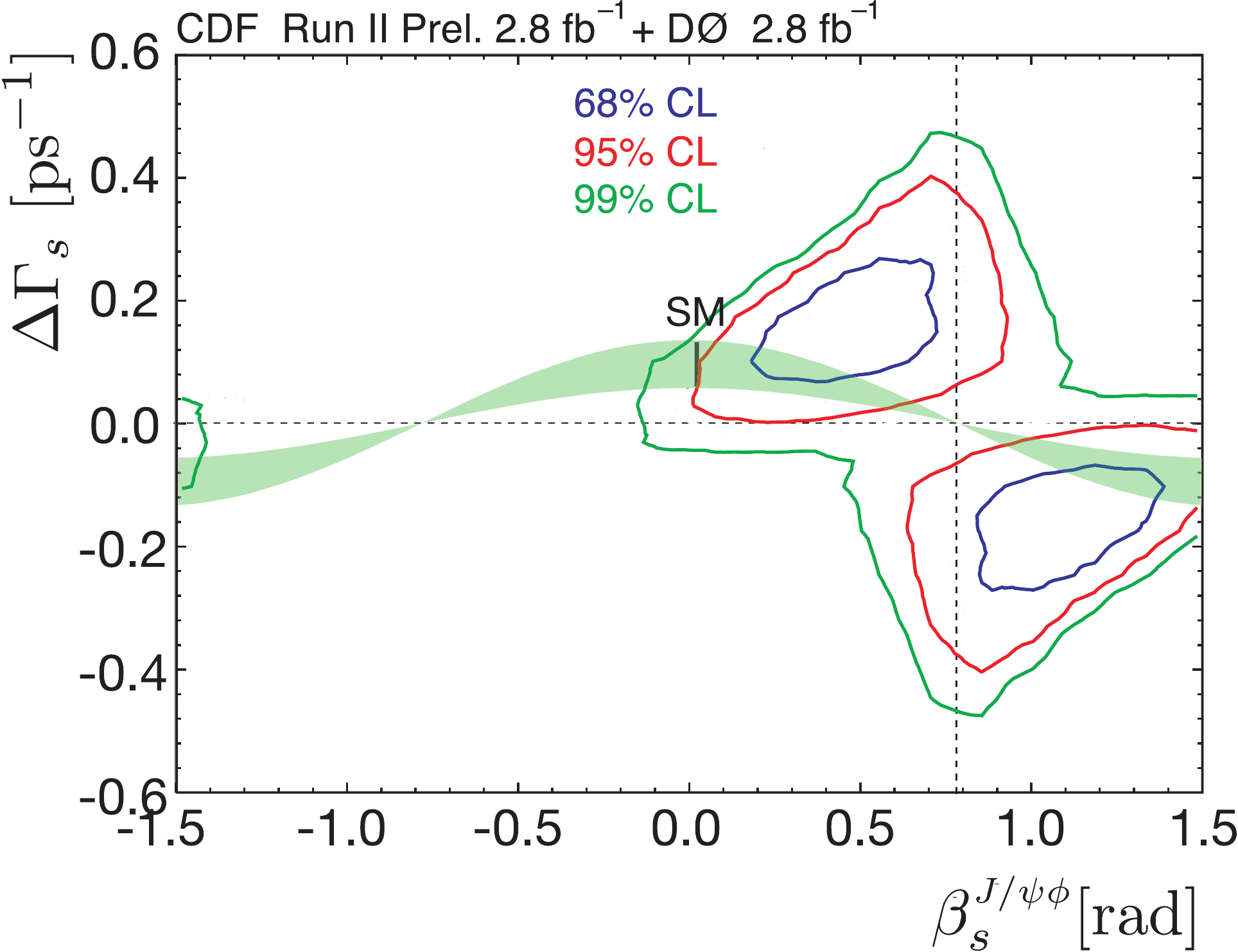}
\hspace{0.1in}
\includegraphics[width=2.35in]{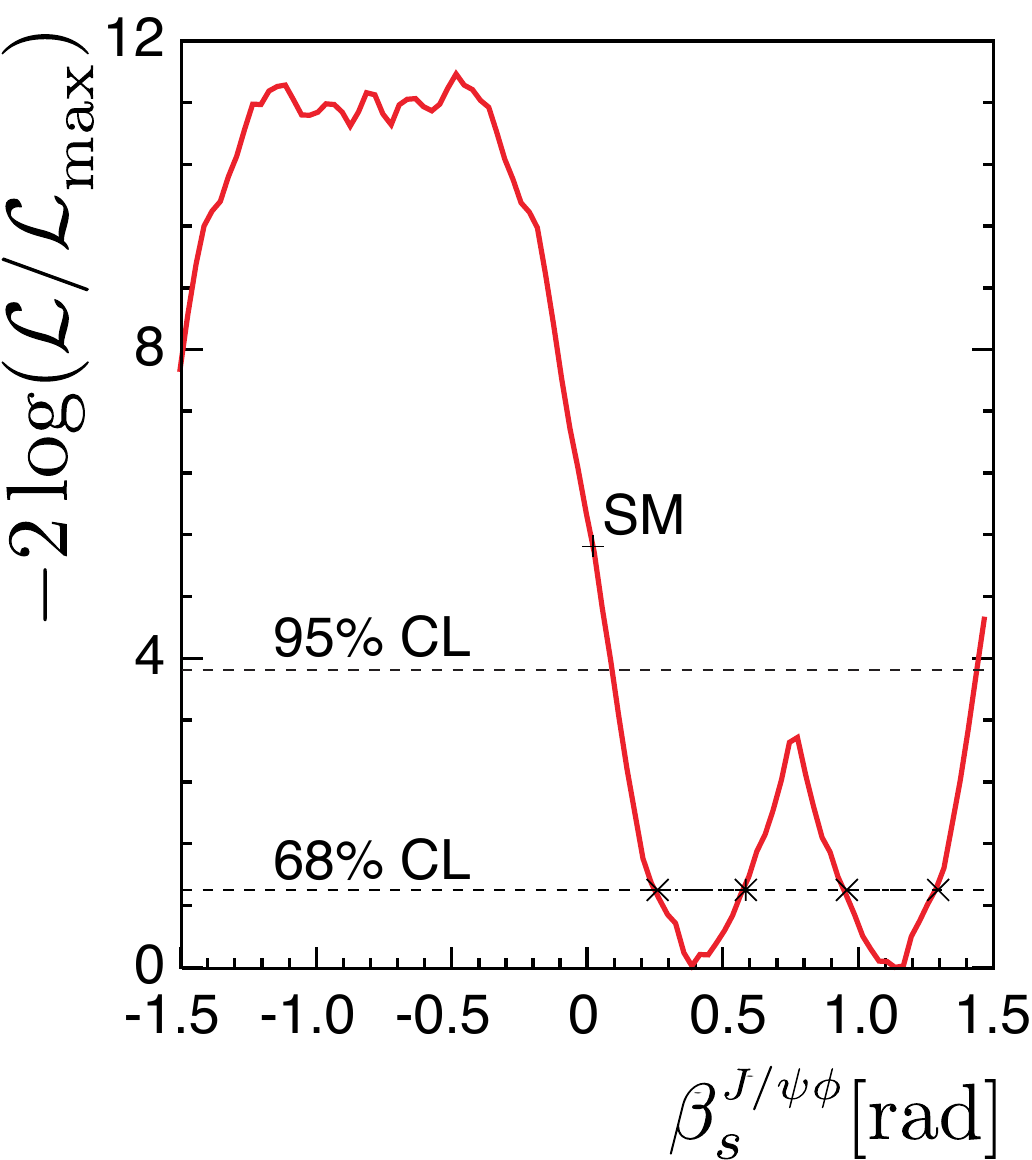}
\caption{(left) Confidence contours for ($\Delta\Gamma_s $, $\mathrm{\phi_s}$) (right) One-dimensional confidence limits on $\beta_s$}
\label{fig:betas_comb}
\end{figure}

The final combined result is shown in Fig.~\ref{fig:betas_comb}.  A 1--dimensional confidence interval for \betas\  is also obtained:

\begin{eqnarray}
\betas\ \in [0.27,0.59] \cup [0.97,1.30] @68\% CL \\
\betas\ \in [0.10,1.42] @95\% CL \\
\end{eqnarray}

The above confidence intervals imply that the SM value $\betas = 2.1\pm 0.7 \times 10^{-3}$ is excluded at more than
 95\% CL. More precisely, a goodness-of-fit test of our data with the SM model , yields a significance of 0.034 (given the SM uncertainty on the value of $\Delta\gamma$, the p-value for the "most conservative" SM point is actually 2.0$\sigma$).
 Comparing this result with the average performed by HFAG 2008 reveals that the level of agreement with SM is essentially unchanged. This comes from the opposite effects of two changes: the increase of statistics on the CDF side, and a better accounting for tails in the \Dzero\ measurement. Further work is in progress to allow a unified simultaneous fit of the CDF and \Dzero\ data samples, which should in principle provide a better resolution because of a more effective use of the information in the parameters that get ``profiled away" in the step of taking 2D projection of each experiment.
 
 In conclusion, the value of \betas\ keeps showing a mild disagreement with the SM expectation. The fit prefers rather large values of \betas , $\approx 0.5$. The jury is still out on whether this is first sign of new physics (many BSM models predict large values of \betas , most notably 4th-generation models) or just a statistical fluctuation. The answer should become clear by the end of run II.
It is interesting to note that in some models, a large phase implies a \Bsmumu  rate just around the corner from current limits.

\subsection{$\bs\to\phi\phi$ }

\begin{figure}
\begin{center}
\includegraphics[width=4in]{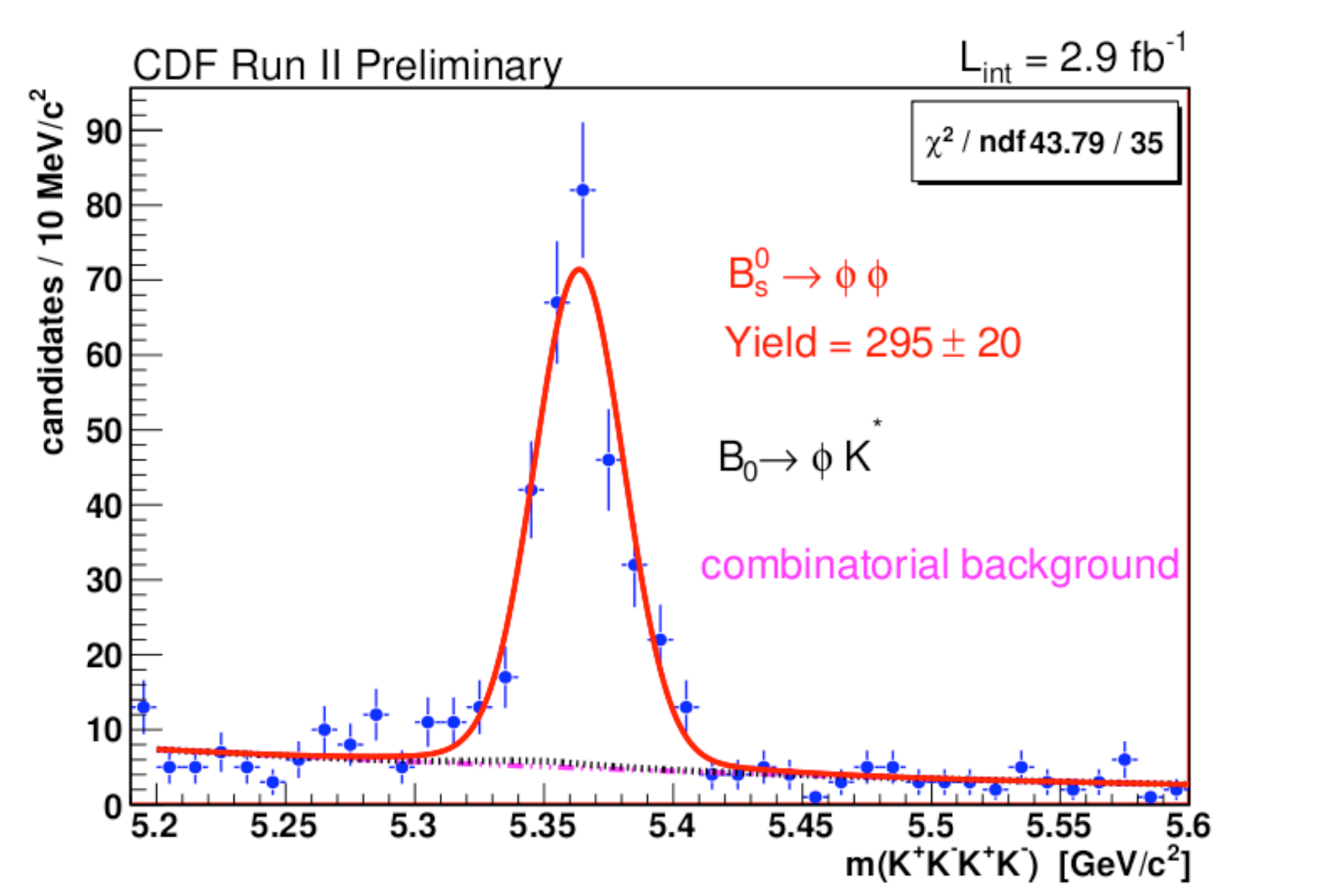}
\caption{$\bs\to\phi\phi$ signal at CDF}
\label{fig:bsphiphi}
\end{center}
\end{figure}

Amongst the most recent results is the high-statistics study of the $\bs\to\phi\phi$ mode. This is a $b\to s\bar{s}s$ pure penguin process and it is sensitive to possible new physics CP violating contributions, both in the decay and in the \bs\ mixing. In addition to A$_{CP}$, its BR and polarization are also good measurements to search for deviations from SM. 

This mode was first observed at CDF in 2005, and still unique to this experiment. 
The first observation was based on 0.2\lumifb\ and only had 8 signal events; the current analysis has exploited not only an increased sample of 2.9\lumifb\ but also an improved acceptance thanks to a better usage of multiple trigger selection, resulting in a sample of about 300 events (Fig.~\ref{fig:bsphiphi}). This allows now a much more accurate determination of BR:
\begin{eqnarray}
\frac{BR(\bs\to\phi\phi)}{BR(\bs\to J/\psi\phi)} = (1.78\pm 0.14\pm 0.20 ) \times 10^{-2}\\
BR(\bs\to\phi\phi) = (2.40\pm 0.21 \pm 0.27 \pm 0.82) \times 10^{-5}
\end{eqnarray}

where the errors are respectively due to statistics, systematics, and branching fraction uncertainty from the PDG. This compares well with predictions of a recent QCDF calculation: $(2.18\pm 0.11(CKM) ^{+3.04}_{-1.7}(th))\times 10^{-5})$~\cite{Beneke:2006hg}, with the precision of the comparison now being dominated by theoretical uncertainties. 

This sample is substantial enough to move to the next step: measurement of polarization amplitudes. This measurement is currently ongoing, and a resolution of ~10\% is expected. This is particularly interesting in the light of current data on other penguin--dominated VV decays, showing  much smaller values for the longitudinal fraction $f_L$ than expected in the SM (``polarization puzzle")~\cite{Bevan}.

\section{Prospects}

At this time, each of the Tevatron experiments has accumulated about 6\lumifb\ of data. Several current results on flavor physics are based on much smaller samples (1 or 2\lumifb ), and many channels have not been explored at all; this is particularly true for the large hadronic samples at CDF. The Tevatron luminosity has been steadily increasing over the past years up to the present date. With the current expectation of running with steady luminosity through year 2011, a further doubling of statistics is expected, reaching a sample of 10-12\lumifb . With this amount of data and as the analysis work progresses, many new results are expected.

\begin{figure}[tb]
\begin{center}
\includegraphics[width=5 in,height=3.5in,clip, trim= 0in 0.5in 0in 0.6in]{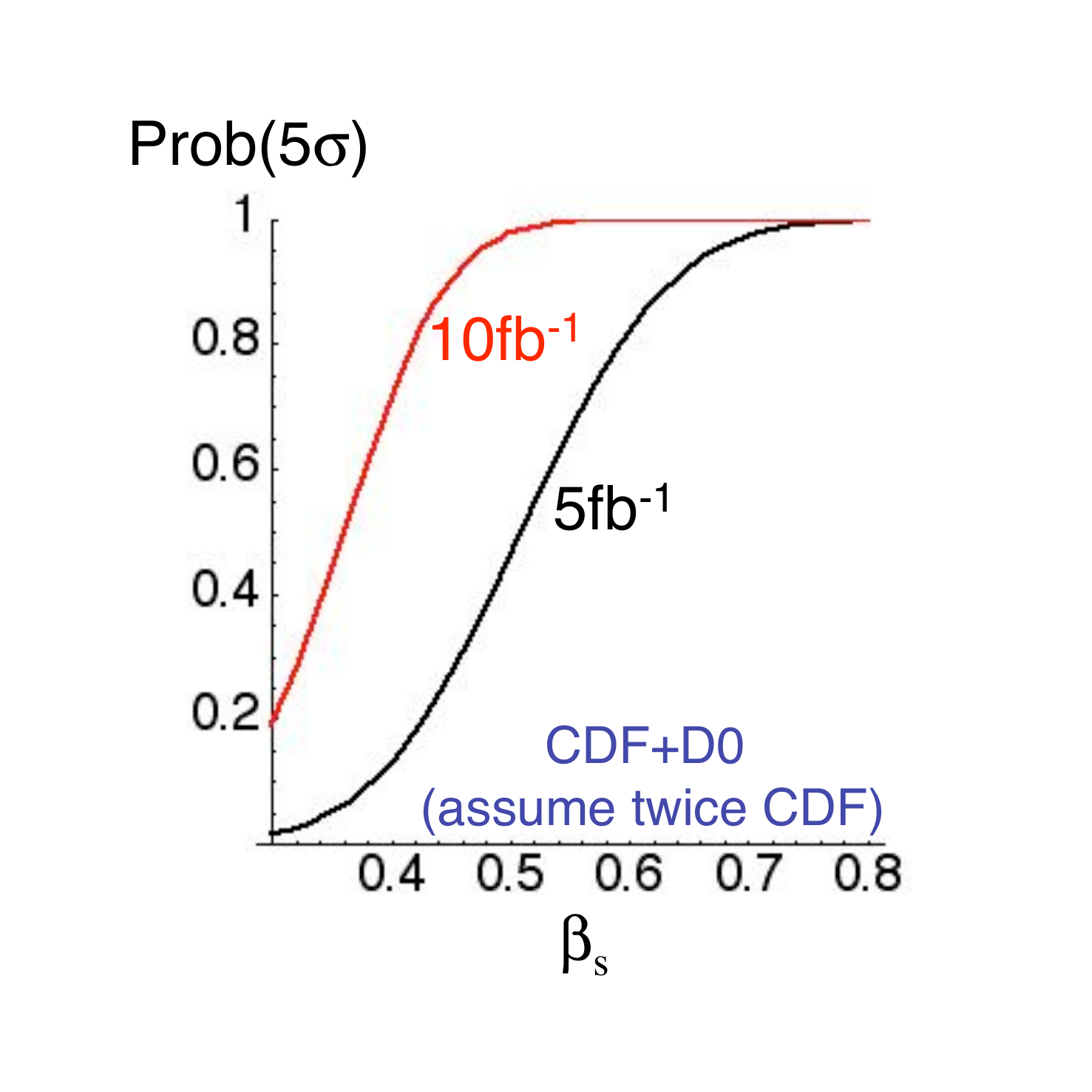}
\caption{Projected probability for observing at the Tevatron a 5$\sigma$ deviation from the SM, as a function of the true value of $\beta_s$}
\label{fig:betas-proj}
\end{center}
\end{figure}

The future sensitivity of the \betas\ analysis has been estimated by CDF, in the conservative assumption of no analysis improvements, and no use of constrains coming from additional measurements like $A_{SL}$ or other external quantities. 
Fig.~\ref{fig:betas-proj} show the probability of observing a 5$\sigma$ deviation from the SM as a function of the true value of \betas , for 10\lumifb\ of collected data, for a single, or the combination of the two experiments (assumed equivalent to CDF). The probability for 5$\sigma$ observation is very high if \betas\ is above 0.4 - 0.5 (the probability of 3$\sigma$ evidence is of course much higher still). It is interesting to note that several BSM physics model do predict values of \betas\ in this range; for instance, 4th generation models have formulated predictions in the range 0.5-0.7~\cite{Hou:2006mx}.
At the same time, the related \bsphiphi\ mode is also going to be studied with larger statistics. Polarization will be measured and eventually CP asymmetry.

\begin{figure}
\includegraphics[width=\textwidth]{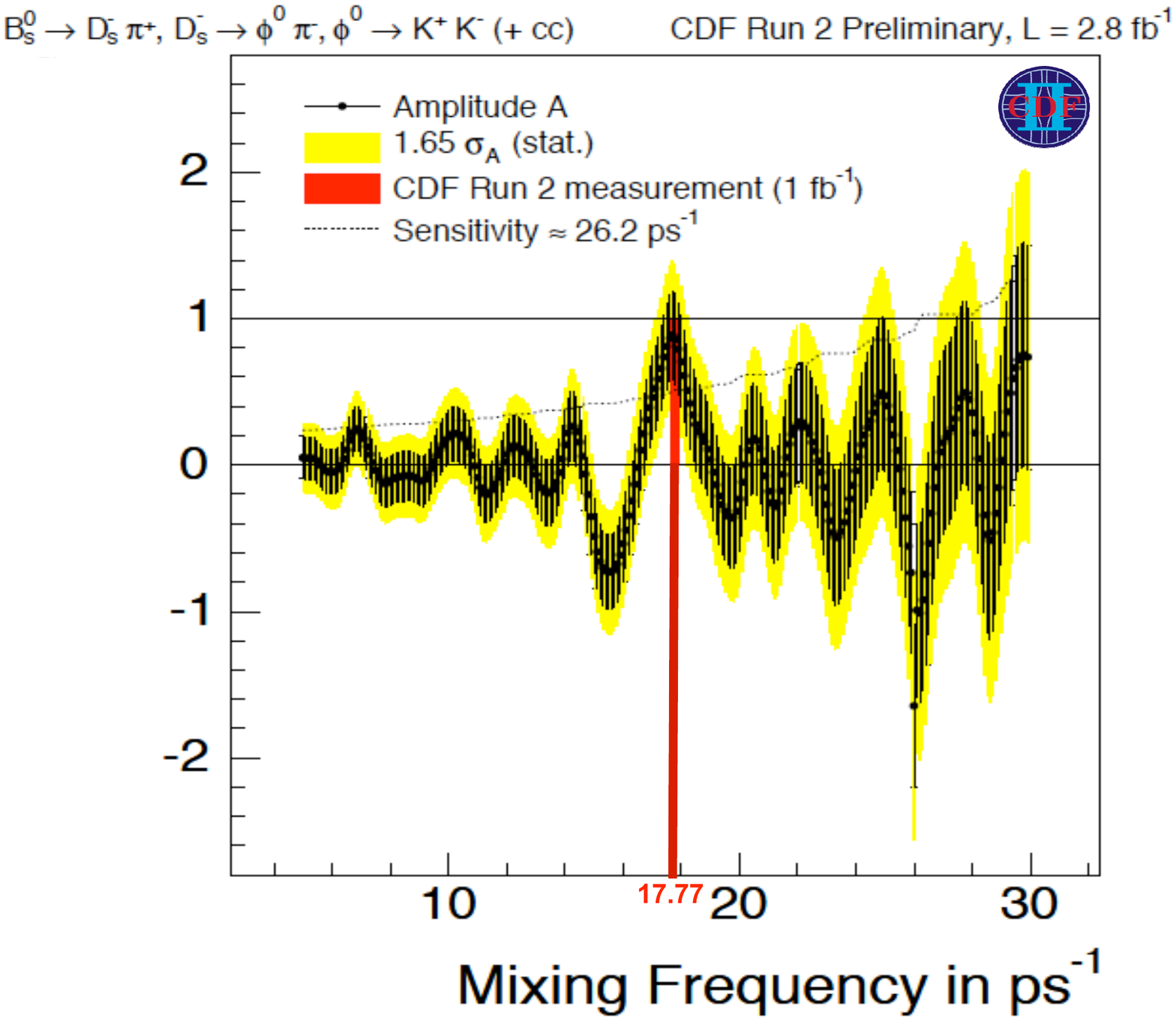}
\caption{Amplitude scan for \bs\  flavor oscillation in the $\bs \to D_s\pi$ mode, based on Same--Side Kaon Tagging.}
\label{fig:bs-osc}
\end{figure}

The increase of statistics brings not only an improvement in the precision of current measurements, but also a significant broadening of the range of measurements that are accessible. The \bs\ oscillations were first observed in 2006 by combining several exclusive and inclusive \bs\ modes, and all available flavor tagging algorithms. Today we can observe a clear oscillation signal in a single fully reconstructed channel, and using a single tagger. Fig.~\ref{fig:bs-osc} show the amplitude scan in 2.8\lumifb\  for $\bs \to D_s\pi$, using only the same--side kaon tagger (SSKT), exhibiting a clear signal at the frequency measured in the original observation~\cite{betas_CDF}. This signal confirms that the tagging capability of CDF has not been hampered by the increased luminosity, and it will allow to calibrate tagger dilution from data, rather than having to rely on Monte Carlo. At the same time, a more powerful flavor tagging algorithm is being developed, using an ANN to combine information from both the same and the opposite side of the \bs\ in a single tagger, to achieve the best possible performance. As a consequence, we are now entering the era of time-dependent CP--asymmetry measurements in the \bs\ system. Examples are the \bskk\ mode, and the determination of angle $\gamma$ from $\bs\to D_s K$. 

Other determinations of $\gamma$ will also become important. CDF has performed a pilot measurement of CP asymmetry in the mode $\bu\to D^0K^+$, where the $D^0$ decays in one of the CP eigenstate modes $D^0\to \pi\pi$ and $D^0\to KK$ ("GLW method"). Using a sample of just 1\lumifb\ it provided a result already of comparable resolution with existing results on the same mode from \ee\ B-factories. Reconstruction of $\bu\to D_{DCS}\pi$ mode was also achieved.
From these results, one can extrapolate that the CDF sample with 10\lumifb\ will be roughly equivalent to 3 ab$^{-1}$ of \ee\ data, which implies it will be the most important measurement at that time, adding on top of the CDF-unique measurements in the $\bs\to D_s K$ mode.

\subsection{Charm physics}

A field that is becoming very important at the Tevatron is charm physics. CDF has a huge sample of $D^0$ decays from its hadronic trigger, and is accumulating further data at a high rate (as an example, the rate of reconstructed $D^0\to K\pi$ events is 4M/year, a rate an order of magnitude larger than the next best experiment, Belle). From these data, measurements of both mixing and CPV have been performed in the past from small samples. The CDF measurement of $D^0$ oscillations in the $K\pi$ channel with a 1.5\lumifb\ sample has produced very similar results to BaBar, both for resolution and values measured for the mixing parameters. The current CDF sample should allow a 5$\sigma$ observation to be performed in a single experiment. 

The measurement of CP violation in $D^0\to \pi\pi$ and $D^0\to KK$ modes published by CDF in 2005 using a sample of just 0.13 \lumifb\ has been the dominant measurement at the time, with a resolution of 1.3\%, and it is still a significant  contribution to today's world average, which has a resolution of 0.4\%~\cite{PDG}. The analysis in progress on the current data sample is expected to improve the world-average precision by a factor of 2, and reach an ultimate precision at the end of the run in the $10^{-3}$ range, a very interesting region for possible new physics effects in the charm system. 

\section*{Acknowledgements}

The author wishes to thank the Fermilab for kind hospitality, and the EPS 2009 organizers for the opportunity to partecipate to a very successful and pleasant conference.

\end{document}